\documentclass[12pt]{article}

\begin{document}

\begin{center} {\Large\bf WKB Method and Relativistic Potential Models}

\bigskip
 V.V. Rubish
\medskip

Department of Theoretical Physics, Uzhhorod National University,
\\ Voloshin str. 32, 88000 Uzhgorod, Ukraine\\ e-mail:
   vrubish@univ.uzhgorod.ua
\end{center}
\bigskip

\begin{abstract}
Based on the Dirac approach we have developed the relativistic
vision of the WKB method for centrally symmetrical potential with
mixed Lorenz structure. We have obtained relativistic
wavefunctions of light quark and the new rule of quantization
containing the spin-orbit interaction. These gives us the
possibility of finding the energy levels and decay width of
hydrogen-like quark-systems.
\bigskip

\end{abstract}

\vspace{0.5cm}\noindent{\Large\bf 1. Introduction}\vspace{0.5cm}

\noindent The Dirac equation with potentials of a concrete type,
for which possible to write the exact solution, meet enough
seldom. Most often for searching the solutions either numerical
or asymptotic methods are used. In many theoretical and applied
problems a possibility of obtaining the asymptotic solution
allows to carry out the fullest analysis of a problem. Therefore
hardly there is a necessity in detail to explain importance of
creating and investigating asymptotic methods of the solving the
Dirac equation.

The Wentzel-Kramers-Brillouin quasi-classical approximation (or
WKB method) is one of basic and most universal asymptotic methods
of solving problems of theoretical and mathematical physics (see,
for example, \cite{Lan, Mig, Head, Fr}), for which the exact
solutions are either unknown or rather onerous. As is known
\cite{Lan, Mig, Head, Fr}, in case of a Coulomb field it has a
high accuracy even for small values of quantum numbers. In
contrast to the perturbation theory the given approach is not
connected with a smallness of interaction and consequently has
more wide applicability area allowing to study qualitative
legitimacies in behaviour and properties of quantum mechanical
systems. The new application of quasi-classical approach can be
low-energy sector of QCD (the energy spectrum of hadrons,
confinement study, decay widths of hadrons), where standard
approaches based on perturbative theories are inapplicable in
consequence of the fact that interaction between quarks in this
area is not small. \vspace{1cm}

\vspace{0.5cm}\noindent{\Large\bf 2. The WKB method for the Dirac
equation in a centrally symmetrical field}\vspace{0.5cm}

\noindent Obtain the formulae of the quasi-classical approximation
for solutions of the Dirac equation with potential of centrally
symmetry having mixed Lorenz structure. Corresponding to this we
search wavefunctions of the stationary states (in standard
representation) in the form of
\begin{equation}
\label{eq1} \Psi=r^{-1}\left(\begin{array}{c}
F(r)\Omega_{jlm}\left({\bf n}\right)\\
i G(r)\Omega_{jl'm}\left({\bf n}\right)
\end{array}\right),
\end{equation}
where $\Omega$ -- the spherical spinor, $j$ and $m$ -- the total
angular moment and projection of $j$ $\left(j=l\pm1/2\right)$, $l$
-- the orbital moment ($l+l'=2j$), ${\bf n}={\bf r}/r$.

After separation the angular variable the Dirac equation with
centrally symmetrical potential containing both vector $V(r)$ and
scalar $S(r)$ parts, takes form ($\hbar=c=1$)
\begin{eqnarray} \frac{dF}{dr}+\frac{k}{r}F-\frac{1}{\hbar}
\left[\left(E-V(r)\right)+\left(m+S(r)\right)\right]G=0, \nonumber \\
\frac{dG}{dr}-\frac{k}{r}G+\frac{1}{\hbar}\left[\left(E-V(r)\right)-\left(m+S(r)\right)\right]F=0,
\label{eq2}
\end{eqnarray}
where $F$ and $G$ are the radial functions, $k=\mp
\left(j+1/2\right)$ for states with $j=l\pm1/2$, $E$ -- the energy
of level.

For finding the quasi-classical solutions of the system
(\ref{eq2}) it is convenient to write equations (\ref{eq2}) in
the matrix form \cite{Mur}:
\begin{eqnarray}
\Psi'=\frac{1}{\hbar}D\Psi,\quad \Psi=\left\{
\begin{array}{c}
F\\G
\end{array} \right\},\quad D=\left(\begin{array}{cc} - \hbar k/r& E-V(r)+m+S(r) \\
-E+V(r)+m+S(r)& \hbar k/r
\end{array}\right). \label{eq3}
\end{eqnarray}
Here a Planck constant $\hbar$ is remained in the explicit view,
the prime denotes the derivative with respect to r. The solution
of the matrix equation (\ref{eq3}) we shall look as the formal
expansion in powers of $\hbar$:
\begin{equation}
\Psi=\varphi \exp\left(\int\limits^{r}y dr\right),\quad
y(r)=\frac{1}{\hbar}y_{-1}(r)+y_{0}(r)+\hbar y_{1}(r)+\hbar^{2}
y_{2}(r)+...,\quad
\varphi(r)=\sum\limits_{n=0}^{\infty}\hbar^{n}\varphi^{(n)}(r),\label{eq4}
\end{equation}
where the upper (lower) component  $\varphi^{(n)}_{F}$ (
$\varphi^{(n)}_{G}$) of the vector $\varphi^{(n)}$ corresponds to
the radial wavefunction $F (G)$. Having substituted (\ref{eq4})
into (\ref{eq3}) and equated to zero the coefficients of each
power of $\hbar$, we arrive at the recurrent system
\begin{eqnarray}
\left(D-y_{-1}\right)\varphi^{(n)}=0,\quad\quad\quad\quad\quad\quad\quad\quad\quad\quad\quad
\quad\quad\quad\nonumber\\
\left(D-y_{-1}\right)\varphi^{(n+1)}=\varphi^{(n)'}+\sum
\limits_{k=0}^{n} y_{n-k}\varphi^{k}, \quad n=0, 1,\ldots
\label{eq5}
\end{eqnarray}
From the first equation of system (\ref{eq5}) follows that $y$ is
eigenvalues, $\varphi^{n}\equiv \varphi_{i}$  is one of (right)
eigenvectors of the matrixes $D$. Eigenvalues
$y_{-1}\left(r\right)\equiv\lambda_{i}$ are roots of the secular
equation $\det\left(D-y_{-1}\right)=0$. Corresponding
eigenvectors $\varphi_{i}$ can be find in explicit form at
diagonalization of the matrix $D-\lambda_{i}$:
\begin{equation}
\label{eq6}
y_{-1}=\pm\sqrt{\left(m+S\right)^{2}-\left(E-V\right)^{2}+\frac{k^{2}}{r^{2}}}=\pm
q=\lambda_{i},
\end{equation}

\begin{equation}
\label{eq7} \varphi_{i}=A_{1}\left(\begin{array}{c} m+S+E-V
\\ kr^{-1}+\lambda_{i}
\end{array}\right)=A_{2}\left(\begin{array}{c}
\lambda_{i}-kr^{-1}\\ m+S-E+V
\end{array}\right).
\end{equation}
Here index $i=\pm$, $A_{1}(r)$ and $A_{2}(r)$ are some functions,
which will be determined below.

Since matrix $D$ is not symmetrical, together with right
eigenvectors $\varphi_{i}$, it is necessary to enter left
eigenvectors $\check{\varphi_{i}}$. They define by conditions
\begin{equation}
\label{eq8} \check{\varphi_{i}}\left(D-\lambda_{i}\right)=0, \quad
\quad \check{\varphi_{i}}\neq \varphi_{i}^{T},
\end{equation}
\begin{eqnarray}
\check{\varphi_{i}}=B_{1}\left(\begin{array}{cc} m+S-E+V,&
kr^{-1}+\lambda_{i}
\end{array}\right)=B_{2}\left(\begin{array}{cc} \lambda_{i}-kr^{-1},&
 m+S+E-V \end{array}\right).\label{eq9}
\end{eqnarray}
Furthermore, left and right eigenvectors the mutually orthogonal
\begin{equation}
\label{eq10} \left(\check{\varphi_{i}},
\varphi_{j}\right)=\sum^{2}_{\alpha=1}\left(\check{\varphi}_{i}\right)_{\alpha}
\left(\varphi_{j}\right)_{\alpha} =\mathrm{const}\cdot\delta_{ij}.
\end{equation}
Remind that $\check{\varphi}_{i}$ there is vector-column, and
$\delta_{ij}$ is the Kronecker symbol. For determination $y_{0}$,
we take $\varphi_{0}=\varphi_{i}$ in the first equation of
(\ref{eq5}) and multiply both parts of this equation by
$\check{\varphi_{i}}$ on the left. Then due to (\ref{eq8}), the
left part is zero and we get the equation for $y_{0}$, from which
\begin{equation}
\label{eq11}
y_{0}(r)=-\frac{\left(\check{\varphi_{i}},\varphi_{i}'\right)}{\left(\check{\varphi_{i}},\varphi_{i}\right)}
\end{equation}

We select functions  $A_{1,2}(r)$, $B_{1,2}(r)$, in (\ref{eq7}),
(\ref{eq9}) so that the condition
 \begin{equation}
\label{eq12}
\left(\check{\varphi_{i}},\varphi_{i}'\right)=\left(\check{\varphi_{i}}',\varphi_{i}\right)
\end{equation}
is realized. Then
\begin{equation}
\label{eq13}
\int\limits^{r}y_{0}dr=\ln\left[\left(\check{\varphi_{i}},\varphi_{i}\right)^{-1/2}\right]
\end{equation}
and as a result we obtain
\begin{equation}
\label{eq14}
\Psi=\varphi_{i}\left(\check{\varphi_{i}},\varphi_{i}\right)^{-1/2}
\exp\left(\frac{1}{\hbar}\int\limits^{r}\lambda_{i}dr\right),
\end{equation}
that is similar usual expression for quasi-classical wavefunction
in nonrelativistic quantum mechanics
\begin{equation}
\label{eq15} \Psi \sim p^{-1/2}\exp\left(\pm i\int\limits^{r}p
dr\right).
\end{equation}

Using the first two equation of system (\ref{eq5}) by means of
left and right vectors technique we find the terms $y_{1}$,
$y_{0}$ and $\varphi^{(0)}$. Solving next equations of the system
(\ref{eq5}) by the similar procedure one can sequentially find
the terms $y_{2}$, $y_{3}$,..., $\varphi^{(2)}$,
$\varphi^{(3)}$,... in the expansions (\ref{eq4}). But formulae
for them are rather cumbersome, therefore in applications ones
usually restrict themselves to only first terms. Actually the
reason of this is the fact  \cite{Lan}, \cite{Mig} that the
expansions in powers of $\hbar$ (\ref{eq4}) in the general case
don't convergent and are asymptotic series, the finite number of
terms of which gives the good approximation for the wavefunction,
if a parameter of an expansion (the Planck constant $\hbar$) is
rather small.

It is always easy to satisfy the condition (\ref{eq12}). By
substituting the expression (\ref{eq7}), (\ref{eq8}), into
(\ref{eq12}) we come to the equation
\begin{equation}
\label{eq16} \frac{A_{1}B_{1}'-A_{1}'B_{1}}{A_{1}B_{1}}=
-\frac{\left(m+S\right)V'+\left(E-V\right)S'}{q\left(q\pm
kr^{-1}\right)},
\end{equation}
whene
\begin{eqnarray}
\label{eq17} \Psi=\left\{
\begin{array}{c}
F\\G
\end{array} \right\}=\left[2q\left(q\pm\frac{k}{r}\right)\right]^{-1/2}
\exp\left\{\pm\int\limits^{r}q dr+\right. \nonumber
\\ \nonumber \\ \left.+\frac{1}{2}\int\limits^{r}\frac{\left(m+S\right)V'+\left(E-V\right)S'}{q\left(q\pm
kr^{-1}\right)}dr\right\} \left(\begin{array}{c}
m+S+E-V \\
kr^{-1}\pm q \end{array}\right)
\end{eqnarray}
(here and hereinafter $\hbar=1$). Using the second form of writing
the eigenvectors $\varphi_{i}$ and $\check{\varphi_{i}}$ (with
multiplier $A_{2}(r)$ and $B_{2}(r)$ in (\ref{eq7}) and
(\ref{eq9})) leads to the following result:
 \begin{eqnarray}
\label{eq18} \Psi=\left\{
\begin{array}{c}
F\\G
\end{array} \right\}=\left[2q\left(q\mp\frac{k}{r}\right)\right]^{-1/2}
\exp\left\{\pm\int\limits^{r}q dr-\right. \nonumber
\\ \nonumber \\
\left.-\frac{1}{2}\int\limits^{r}\frac{\left(m+S\right)V'+\left(E-V\right)S'}{q\left(q\mp
kr^{-1}\right)}dr\right\} \left(\begin{array}{c}
\pm q -kr^{-1}\\
m+S-E+V \end{array}\right).
\end{eqnarray}
We note that $q$ coincides (to within multiplier $i=\sqrt{-1}$)
with radial momentum of the relativistic quasi-classical
particle. The sign $+ (-)$ in (\ref{eq17}) and (\ref{eq18})
corresponds to the solution increasing (decreasing) with
increasing $r$. For the decreasing solutions (the sign $-$) it is
necessary to use the formula (\ref{eq17}) when $k<0$ and formula
(\ref{eq18}) when $k>0$; for increasing solution -- on the
contrary. The choice of the suitable form of writing the solution
is defined by that to value $Q=q+|k|/r$ should be positive in
below-barrier region.

The effective potential
\begin{equation}
\label{eq19}U\left(r,E\right)=\frac{E}{m}V+S+\frac{S^{2}-V^{2}}{2m}+\frac{k^{2}}{2mr^{2}}
\end{equation}
corresponds to the Dirac system (\ref{eq2}). So
$U\left(r,E\right)$ looks like a potential with a barrier. We
consider the most general case, when the effective potential is
barrier type. Then wavefunction has the different view in the
three regions: 1) potential well $r_{0}<r<r_{-}$ ($q^{2}<0$); 2)
the below-barrier region $r_{-}<r<r_{+}$ ($q^{2}>0$); 3) the
classically allowed region with continuum spectrum $r>r_{+}$ (
$q^{2}<0$); 4) classically forbidden region $r<r_{0}$
($q^{2}>0$), where $r_{0}$, $r_{-}$ and $r_{+}$ are turning
points. In following section we consider behaviour of the
solutions in these regions.

\vspace{0.5cm}\noindent{\Large\bf 3. The wavefunction of the Dirac
particle in classically allowed and forbidden
regions}\vspace{0.5cm}

\noindent The wavefunction of state has the different form in the
various regions.

\noindent I. The region $r_{0}<r<r_{-}$ is classically allowed;
there the wavefunctions (\ref{eq17}), (\ref{eq18}) oscillate
\begin{equation}
F=C_{1}^{\pm}\left[\frac{E-V+m+S}{p}\right]^{1/2}\cos\Theta_{1},\quad
G=C_{1}^{\pm}\mathrm{sgn}k
\left[\frac{E-V-m-S}{p}\right]^{1/2}\cos\Theta_{2}. \label{eq20}
\end{equation}
where
\begin{equation} \label{eq21}
p(r)=\sqrt{\left(E-V\right)^{2}-\left(m+S\right)^{2}-\left(k/r\right)^{2}}
\end{equation}
is the quasi-classical momentum for the radial motion of a
particle, $C_{1}^{\pm}$ -- normalization constant.
\begin{equation}\label{eq22}
\Theta_{1}=\int\limits
_{r_{-}}^{r}\left(p+\frac{kw}{pr}\right)dr+\frac{\pi}{4},\quad
\Theta_{2}=\int\limits
_{r_{-}}^{r}\left(p+\frac{k\widetilde{w}}{pr}\right)dr+\frac{\pi}{4}.
\end{equation}
\[
w=\frac{1}{2}\left(\frac{V'-S'}{m+S+E-V}-\frac{1}{r}\right),\quad\quad
\widetilde{w}=\frac{1}{2}\left(\frac{V'+S'}{m+S-E+V}+\frac{1}{r}\right).
\]
Signs $\pm$ correspond to values $k>0$ and $k<0$. If the level
width $\Gamma$ is small (it will be shown later) the wavefunction
of quasi-stationary state can be normalized on a single particle
localized in the region I, neglecting its penetrability into the
classically forbidden regions $r<r_{0}$ and $r>r_{-}$ \cite{Lan}
\begin{equation}\label{eq23}
\int \limits_{r_{0}}^{r_{-}}\left(F^{2}+G^{2}\right)dr=1.
\end{equation}
Here $\cos^{2}\Theta_{i}\left(r\right)$ can be replaced with
average value 1/2:
\begin{equation}\label{eq24}
\left|C_{1}^{\pm}\right|=\left\{\int\limits_{r_{0}}^{r_{-}}
\frac{E-V\left(r\right)}{p\left(r\right)}dr\right\}^{-1/2}=\left(\frac{2}{T}\right)^{1/2},
\end{equation}
where $T$-- the frequency period of a relativistic particle
inside a potential well. We note that in turning points $r_{0}$
and $r_{-}$ equation
\[
E-V=\left[\left(m+S\right)^{2}+k/r^{2}\right]^{1/2}
\]
is true and $E-V>m+S$ in region I.

\noindent II. The below-barrier region $r_{-}<r<r_{+}$ is
classically forbidden. Here $p=iq$, and quantities $q$, $y_{-1}$
and $y_{0}$ are real. As known \cite{Lan} the wavefunction should
exponentially decreases inside of this region. So the solutions
of the Dirac system (\ref{eq2}) in the below-barrier region

\noindent for $k<0$ are
\begin{eqnarray}
\label{eq25} \Psi= \frac{C_{2}^{+}}{\sqrt{qQ}} \exp
\left[\int\limits_{r_{+}}^{r}\left(-q
-\frac{\left(m+S\right)V'+\left(E-V\right)S'}{2qQ}\right)dr\right]
\left(\begin{array}{c} -Q\\ m+S-E+V
\end{array}\right),
\end{eqnarray}
for $k>0$
\begin{eqnarray}
\label{eq26} \Psi= \frac{C_{2}^{-}}{\sqrt{qQ}} \exp
\left[\int\limits_{r_{+}}^{r}\left(-q
+\frac{\left(m+S\right)V'+\left(E-V\right)S'}{2qQ}\right)dr\right]
\left(\begin{array}{c} m+S+E-V \\ -Q
\end{array}\right).
\end{eqnarray}

\noindent III. The result for region of continuum ($r>r_{+}$) is
the most interesting. Here the wavefunction corresponds to the
divergent wave (taking off particle):

\noindent for $k>0$
\begin{eqnarray}
\label{eq27} \Psi= \frac{C_{3}^{+}}{\sqrt{pP}} \exp
\left[\int\limits_{r_{+}}^{r}\left(ip
+\frac{\left(m+S\right)V'+\left(E-V\right)S'}{2pP}\right)dr\right]
\left(\begin{array}{c} iP\\ m+S-E+V
\end{array}\right),
\end{eqnarray}
for $k<0$
\begin{eqnarray}
\label{eq28} \Psi= \frac{C_{3}^{-}}{\sqrt{pP}} \exp
\left[\int\limits_{r_{+}}^{r}\left(ip
-\frac{\left(m+S\right)V'+\left(E-V\right)S'}{2pP}\right)dr\right]
\left(\begin{array}{c} m+S+E-V \\ iP
\end{array}\right),
\end{eqnarray}
where $P=p+i\left|k\right|/r$.

\noindent IV. In classically forbidden region ($r<r_{0}$) the
wavefunctions are of the form:

\noindent for $k<0$
\begin{eqnarray}
\label{eq29} \Psi= \frac{C_{4}^{-}}{\sqrt{q\left(q-k/r\right)}}
\exp \left[\int\limits_{r_{0}}^{r}\left(q
+\frac{\left(m+S\right)V'+\left(E-V\right)S'}{2q\left(q-k/r\right)}\right)dr\right]
\left(\begin{array}{c} E-V+m+S \\ q-k/r
\end{array}\right),
\end{eqnarray}
for $k>0$
\begin{eqnarray}
\label{eq30} \Psi= \frac{C_{4}^{+}}{\sqrt{q\left(q+k/r\right)}}
\exp \left[\int\limits_{r_{0}}^{r}\left(q
-\frac{\left(m+S\right)V'+\left(E-V\right)S'}{2q\left(q+k/r\right)}\right)dr\right]
\left(\begin{array}{c} q+k/r\\ m+S-E+V
\end{array}\right).
\end{eqnarray}

The formulae (\ref{eq20})-(\ref{eq30}) include the whole range of
values of $r$, except for neighborhoods of turning points. For
bypass of these points and matching the solutions we use the
usual method \cite{Mig}. Closely to the $r_{-}$ and $r_{+}$ the
system (\ref{eq2}) reduces to the Schr\"{o}dinger equation with
the effective potential linearly depending on $r-r_{\pm}$, the
solution of which expressed through the Airy function; one can
match by the more elegant Zwaan method \cite{Lan}, \cite{Head}. So
the relation between the constants in various regions is of the
form
\begin{eqnarray}
C_{2}^{\pm}=i
C_{3}^{\pm}=\quad\quad\quad\quad\quad\quad\quad\quad\quad\quad\quad\quad\quad\quad\quad\quad\quad\quad\nonumber
\\ \mp\frac{C_{1}^{\pm}}{2}
\left[\frac{E-V\left(r_{-}\right)+m+S\left(r_{-}\right)}
{\left|k\right|r_{-}^{-1}}\right]^{\pm\frac{1}{2}}\exp
\left[-\int\limits_{r_{-}}^{r_{+}}\left(q \pm
\frac{\left(m+S\right)V'+\left(E-V\right)S'}{2qQ}\right)dr\right],\nonumber\\
2C_{4}^{\pm}\left[\frac{\left|k\right|r_{0}^{-1}}{E-V\left(r_{0}\right)
+m+S\left(r_{0}\right)}\right]^{\frac{\mathrm{sgn}k}{2}}=\left(-1\right)^{n}C_{1}^{\pm}.\label{eq31}
\quad\quad\quad\quad\quad\quad
\end{eqnarray}

Let us find the equation determining the real part of the level
energy $E$ and width $\Gamma$ of quasi-stationary levels
$E=E_{r}-i\Gamma/2$. Neglecting the barrier penetrability from
(\ref{eq23}) we obtain the quantization condition:
\begin{equation} \label{eq32}
\int\limits_{r_{0}}^{r_{-}}\left(p+\frac{kw}{pr}\right)dr
=\left(n+\frac{1}{2}\right)\pi,
\end{equation}
where $n=0,1,2,\ldots$ is the radial quantum number.

Pass to calculation of the level width
\begin{equation}\label{eq33}
\Gamma=-2\mathrm{Im}\left[G^{\ast}F\right]_{r\rightarrow\infty}.
\end{equation}
Having used the explicit expressions (\ref{eq27}), (\ref{eq28})
for $F$ and $G$ functions, relation between normalization
constants (\ref{eq31}) and expression for $C^{\pm}_{1}$
(\ref{eq24}), we get the tunnelling probability:
\begin{equation}\label{eq34}
\Gamma=\frac{1}{T}\exp\left[-2\int\limits_{r_{-}}^{r_{+}}
\left(q-\frac{kw}{qr}\right)dr\right].
\end{equation}
Though the formulae (\ref{eq23})-(\ref{eq34}) essentially differ
from the formulae of nonrelativistic quasi-classics (in
particular, by relativistic expression for quasimomentum $p$,
taking into account spin-orbit interaction, and additional
pre-exponent multiplier) and more complicated from them. Their
application to concrete problems does not meet difficulties,
since all quantities in functions $F$ and $G$ are expressed in
quadratures.

\vspace{0.5cm}\noindent{\Large\bf 4. Approbation of results
obtained}\vspace{0.5cm}

\noindent{\bf 4.1.} For testing elaborated by us version of the
WKB method, get the known in atomic physics the Bohr-Sommerfeld
formula. We choose $V(r)=-\frac{\alpha Z}{r}$ and $S(r)=0$
($\alpha\approx 1/137$ -- the fine structure constant). Then the
formulae (\ref{eq21}), (\ref{eq22}) are of the form
\[
p(r)=\left[\left(E+\frac{\alpha
Z}{r}\right)^{2}-m^{2}-\left(k/r\right)^{2}\right]^{1/2},
\quad\quad w=\frac{1}{2}\left[\frac{\alpha
Z/r^{2}}{E+\frac{\alpha Z}{r}+m}-\frac{1}{r}\right].
\]
Having calculated the integral from $r_{0}$ to $r_{-}$
\[
r_{0,-}=\frac{E\alpha Z\mp\sqrt{(E\alpha
Z)^{2}-\left(m^{2}-E^{2}\right) \left(k^{2}-\left(\alpha
Z\right)^{2}\right)}}{m^{2}-E^{2}}
\]
in quantization condition (\ref{eq32}), we arrive at the
expression
\[
\frac{E\alpha Z}{\sqrt{m^{2}-E^{2}}}=n+\sqrt{k^{2}-\left(\alpha
Z\right)^{2}},
\]
whence we get expression, which is similar to the Bohr-Sommerfeld
formula
\[
E=\pm m \left[1+\left(\frac{\alpha Z}{n+\sqrt{k^{2}-\left(\alpha Z
\right)^{2}}}\right)^{2}\right]^{-1/2}.
\]

{\bf 4.2.} One more rather interesting case, in which it is
possible to get exact solution of the Dirac equation with
potential of the type $V(r)=-\frac{\alpha}{r}$,
$S(r)=-\frac{\alpha'}{r}$, exists. In the same way that the
Coulomb potential $V(r)$ is derived from the exchange of massless
photon between the nucleus and the leptons orbiting around it, the
scalar potential $S(r)$ is created by the exchange of massless
scalar mesons. The $\sigma$ meson frequently quoted in the
literature has a very high mass and therefore the corresponding
potential has a very short range. In this case the formulae
(\ref{eq21}), (\ref{eq22}) take the form
\[
p(r)=\left[\left(E+\frac{\alpha}{r}\right)^{2}-\left(m-\frac{\alpha'}{r}\right)^{2}
-\left(k/r\right)^{2}\right]^{1/2}, \quad \quad w=
\frac{1}{2}\left[\frac{(\alpha-\alpha')
/r^{2}}{E+\frac{\alpha-\alpha'}{r}+m}-\frac{1}{r}\right].
\]
Quantization condition (\ref{eq32}) gives following result
\[
\frac{E\alpha+m\alpha'}{\sqrt{m^{2}-E^{2}}}=n+\gamma,\quad\quad
\gamma=\pm\sqrt{k^{2}-\alpha^{2}+\alpha'^{2}},
\]
there was the integration from $r_{0}$ to $r_{-}$
\[
r_{0,-}=\frac{E\alpha+m \alpha' \mp\sqrt{(E\alpha+m
\alpha')^{2}-\left(m^{2}-E^{2}\right) \gamma^{2}}}{m^{2}-E^{2}}.
\]
Thus we find the formula coinciding with the expression obtained
in \cite{Gr}:
\[
E=m\left\{\frac{-\alpha\alpha'}{\alpha^{2}+\left(n+\gamma\right)^{2}}
\mp\left[\left(\frac{\alpha\alpha'}{\alpha^{2}+\left(n+\gamma\right)^{2}}\right)^{2}
-\frac{\alpha'^{2}-\left(n+\gamma\right)^{2}}
{\alpha^{2}+\left(n+\gamma\right)^{2}}\right]^{1/2}\right\}.
\]

{\bf 4.3.} We consider the potential
$V\left(r\right)=S\left(r\right)=ar^{2}/4$, $a$ is the force of an
oscillator. In this case the formulae (\ref{eq21}), (\ref{eq22})
is given by
\[
p(r)=\left[E^{2}-m^{2}-\left(E+m\right)\frac{a}{2}r^{2}-\left(k/r\right)^{2}\right]^{1/2},
\quad\quad w= -\frac{1}{2r}.
\]
We integrate in quantization condition (32) from $r_{0}$ to
$r_{-}$
\[
r_{0,-}=\frac{1}{\sqrt{a}}\left[\left(E-m \mp
\sqrt{\left(E-m\right)^{2}-\frac{2ak^{2}}{E+m}}\right)\right]^{1/2}
\]
and gives following result
\[
(E-m)\sqrt{\frac{E+m}{2a}}-|k|+\frac{1}{2}\mathrm{sgn}k=2n,
\]
for $k<0$ get
\[
 \left(2\left|k\right|+1\right)\sqrt{a}-\left(E-m\right)\sqrt{2\left(E+m\right)}
 =-4n\sqrt{a}.
 \]
The last equation for energy is cubic, its real solution is
\[
E_{n,k}=\left(2m+8\cdot2^{2/3}m^{2}A^{-1/3}+2^{2/3}A^{1/3}\right)/6,
\]
where $A=-B+\sqrt{B^{2}-1024m^{6}}$,
$B=32m^{3}-27a\left(1+2k+4n\right)$. This expression completely
coincides with result obtained in \cite{Qia}.

\vspace{0.5cm}\noindent{\Large\bf 5. Description of the energy
spectrum in two-quarks systems with Cornell
potential}\vspace{0.5cm}

\noindent Relativistic description of the bounded states always
was an important problem of nuclear physics and elementary
particle physics, in which relativistic characteristics of the
light quarks play the important role. Our world mainly consists
of the light $u$ and $d$ quarks (the protons - $u\bar{u}d$ and
neutrons - $ud\bar{d}$).

Quantum chromodynamics (QCD) based on principles of the quantum
field theory is justly considered as the most consequent approach
to solving this problem. However, standard perturbative QCD gives
rather reliable recipes of the calculation of various
characteristics only for description so named "hard" processes
which are characterized by large transmitted momentum, and not
applicable for calculation of the characteristics which are
defined by "soft" processes (the mass spectrum, confinement of
quarks, decay widths of hadrons). In the same time
nonperturbative effects basically define the nature of forming
bound states of interacting particles. Confinement is the result
of circumstance that, unlike quantum electrodynamics in which
interaction mediators - photons - are electroneutral, exchange
particles in quantum chromodynamics - gluons - possess non-zero
colour charge and therefore can interact one with other. Thus
confinement is not nested in the framework of the perturbation
theory. On account of this in the given time from principles of
quantum chromodynamics the  structure of interquark forces cannot
completely be defined.

Just study of relativistic properties of quarks systems, namely
of effects caused by spins of quarks, enables to improve  both
the form of confinement part of potential, and Lorenz structure
of potential.

In the work \cite{Pop} starting from the QCD Lagrangian and taking
into account both perturbative and nonperturbative effects, using
the method of vacuum correlators the Dirac equation is obtained
(rigorously for the Coulomb interaction and heuristically for the
confining potential) for a system consisting of a heavy antiquark
(quark) $Q$ ($b$ and $c$ quarks) with the mass $m_{2}$ and light
quark (antiquark) $q$ ($u$, $d$, $s$ - quarks) with mass much less
$m_{1}$.

Such quark systems ($B^{+}=\overline{b}u$, $B^{0}=\overline{b}d$,
$B^{0}_{s}=\overline{b}s$, $D^{0}=\overline{c}u$,
$D^{-}=\overline{c}d$, $D^{-}_{s}=\overline{c}s$) are QCD
analogues of the hydrogen atom and thus its study is of
fundamental importance. From the theoretical viewpoint the
interest in heavy-light systems stems from several
considerations.

First, in the limit of one infinitely heavy quark, one hopes to
get the dynamics of a light quark in the external field of a
heavy one. That would be similar to the picture of the hydrogen
atom.

Second, since the external field is time-independent, one may
hope to obtain a static potential in QCD together with
spin-dependent forces, as has been done for heavy quarkonia.

Third, in the heavy-light system one may study how the chiral
symmetry breaking (CSB) affects the spectrum. When one quark mass
is vanishing, in the chiral symmetric case the spectrum would
consist of parity doublets, and the CSB would lift the
degeneracy.

Fourth, using the Dirac equation we implement explicitly
relativistic dynamics and can study relativistic properties of
the spectrum, e.g., in the case of a vanishing quark mass, and
also relativistic spin splitting in the spectrum.

In the work \cite{Pop} it is explicitly shown that a reasonable
spectrum in Dirac equation occurs only when the confining
potential is a Lorentz scalar and the Coulomb potential is the
fourth component of a 4-vector. In this case the scalar potential
breaks explicitly chiral symmetry, and states with opposite
parities are not degenerate. In the case of a vector confining
part, only quasi-stationary states are found.

We consider the most general type of the confining part, which
taking into account also cases considered in \cite{Pop}. Let us
suggest that the static quark potential has vector and scalar
properties of Lorenz transformation:
\begin{equation}
V_{\mathrm{NR}}\left( r\right) =V_{v}\left( r\right) +V_{s}\left(
r\right) . \label{eq35}
\end{equation}
The quantity $V$ means that the potential is a 4-th component of
operator $\widehat{p}_{\mu}$, $S$ means that the potential has
scalar nature.

Following many authors \cite{Fau, Len} we assume the mixture of
vector-scalar quark interaction potential
\begin{equation}
V\left( r\right) =V_{\mathrm{OGE}}+\varepsilon
V_{\mathrm{conf}},\quad\quad  V_{s}\left( r\right) =(1-\varepsilon
)V_{\mathrm{conf}}, \label{eq36}
\end{equation}
where $V_{\mathrm{OGE}}$ is the one-gluon exchange (OGE) term,
$V_{\mathrm{conf}}$ is the confinement part of the potential,
$\varepsilon $ is mixing constant. Here the Lorenz nature of the
one-gluon part of potential and the confining potential is
different, the one-gluon potential is totally vector, while the
confinement is a vector-scalar mixture. Very interesting reviews
were done in \cite{Br, Sim} concerning the choice of interaction
potential.

To simplify the calculations we consider the simplest case when
$\varepsilon=0$. This corresponds to the one-gluon potential being
totally vector and the confinement potential being totally
scalar.

So we choose the scalar and vector parts of the potential
(\ref{eq36}) in the form \cite{Fau, Len, Br, Sim}
\begin{equation}
V_{\mathrm{OGE}}\left( r\right)=-\frac{\beta}{r}, \quad\quad
V_{\mathrm{conf}}\left( r\right)=\mu r \label{eq37}
\end{equation}
where $\beta=\frac{4}{3}\alpha_{s}$, $\mu=0.18$ GeV$^{2}$,
$\alpha_{s}$ is the constant of the strong coupling with
asymptotic freedom:
\begin{equation}
\alpha _{s}\left( r\right) =\frac{12\pi }{33-2N}\cdot
\frac{1}{\ln \left(Q / \Lambda \right) ^{2}}, \label{eq38}
\end{equation}
where $Q$ is the transmitted momentum, $\Lambda$ is taken to be
equal to $\Lambda=0.14$ GeV, $N$ is the quarks flavours number.

In work \cite{Seet} the WKB method was used for potential
(\ref{eq37}) in the Schr\"{o}dinger equation. For solving of the
Dirac equation system (\ref{eq2}) with potential (\ref{eq37}), we
shall use early elaborated by us version of the WKB approximation.

Due to the confinement of quarks we are interested in only
classically allowed region ($r_{0}<r<r_{-}$, $q^{2}<0$, a
potential well) that corresponds to only the discrete energy
spectrum of the quark-antiquark system. Thus the quasi-classical
momentum (\ref{eq21}) is of the form ($\hbar=c=1$):
\begin{equation}
\label{eq39} p\left(r\right)=\left[\left(E+\frac{\beta}{r}
\right)^{2}-\left(m+\mu
r\right)^{2}-\left(k/r\right)^{2}\right]^{1/2}.
\end{equation}
We represent left part of the quantization condition (\ref{eq32})
in the form of sum of two integrals $I_{1}$ and $I_{2}$
\begin{equation} \label{eq40}
I_{1}=\int\limits_{b}^{a}p(r)dr,\quad\quad\quad
I_{2}=\int\limits_{b}^{a}\frac{kw}{pr}dr.
\end{equation}
The integration is between the two classical turning points
$r_{0}=b$ and $r_{-}=a$, which are real and positive roots
($a>b>0$) of the equation
\begin{equation}\label{eq41}
r^{4}+\frac{2m}{\mu}r^{3}-\frac{E^{2}-m^{2}}{\mu^{2}}r^{2}-\frac{2E\beta}{\mu^{2}}r+
\frac{k^{2}-\beta^{2}}{\mu^{2}}=0.
\end{equation}
Two roots of this equation are real and negatives ($d<c<0$).
Formulae (\ref{eq40}) can now be re-expressed in terms of $a$,
$b$, $c$ and $d$ as
\begin{equation}\label{eq42}
I_{1}=-\mu
\int\limits_{b}^{a}\frac{\left(r^{3}+\frac{2m}{\mu}r^{2}-\frac{E^{2}-m^{2}}{\mu^{2}}r-
\frac{2E\beta}{\mu^{2}}+\frac{k^{2}-\beta^{2}}{\mu^{2}r}\right)}{\left[
\left(a-r\right)\left(r-b\right)\left(r-c\right)\left(r-d\right)\right]^{1/2}}dr
\end{equation}
and
\begin{eqnarray}
I_{2}=-\frac{k}{\mu \left(A-B\right)}
\left[\left(A+\frac{E+m}{2\mu}\right)\int\limits_{b}^{a}
\frac{dr}{\left(r-A\right)\left[
\left(a-r\right)\left(r-b\right)\left(r-c\right)\left(r-d\right)\right]^{1/2}}-\right.
\nonumber \\
\left.-\left(B+\frac{E+m}{2\mu}\right)\int\limits_{b}^{a}
\frac{dr}{\left(r-B\right)\left[
\left(a-r\right)\left(r-b\right)\left(r-c\right)\left(r-d\right)\right]^{1/2}}\right],
\label{eq43}
\end{eqnarray}
where $A$ and $B$, which are roots of the quadratic equation $\mu
r^{2}+\left(E+m\right)r+\beta=0$. These equations can be
represented in terms of complete elliptic integrals. Then
quatization condition (\ref{eq32}) gives following result
\begin{eqnarray}
-\frac{2}{\sqrt{\left(a-c\right)\left(b-d\right)}}\left[\frac{\left(b-c\right)^{2}}
{\left(1-\nu\right)\left(\chi^{2}-\nu\right)}\left[N_{1}\cdot
F\left(\chi\right)+N_{2}\cdot E\left(\chi\right)+N_{3}\cdot
\Pi\left(\nu,\chi\right)+\right.\right.\nonumber\\
+\left.N_{4}\cdot
\Pi\left(\frac{c}{b}\nu,\chi\right)\right]+\frac{k}{\mu}\left.\left[\frac{b-c}{A-B}\left(N_{5}\cdot
\Pi\left(\nu_{1},\chi\right)-N_{6}\cdot
\Pi\left(\nu_{1},\chi\right)\right)+N_{7}\cdot
F\left(\chi\right)\right]\right]= \nonumber \\
=\left(n+\frac{1}{2}\right)\pi, \label{eq44}
\end{eqnarray}
where $F\left(\chi\right)$, $E\left(\chi\right)$ and
$\Pi\left(\nu,\chi\right)$ are complete elliptic integrals of the
first, second and third kinds and
\begin{eqnarray}
N_{1}=\left[\frac{\chi^{2}\mu
\left(b-c\right)}{4}-
\frac{3\mu\left(\nu^{2}-2\nu-2\nu\chi^{2}+3\chi^{2}\right)
\left(b-c\right)}{8\left(1-\nu\right)}-\left(m+\frac{3}{2}c\mu\right)\left(\chi^{2}-\nu\right)+\quad\quad\quad\quad\right.\nonumber\\
\left.+\frac{\left(1-\nu\right)\left(\chi^{2}-\nu\right)}{\left(b-c\right)^{2}}
\left(2c^{2}m +c^{3}\mu+\frac{k^{2}-\beta^{2}}{c\mu}-\frac{2\beta
E}{\mu}-\frac{c\left(E^{2}-m^{2}\right)}{\mu}\right)\right],\quad\quad\quad\quad\quad\quad\quad\nonumber
\end{eqnarray}
\begin{eqnarray}
N_{2}=-\nu\left[m+\frac{3}{2}c\mu+\frac{3}{8}\frac{\mu\left(\nu^{2}-2\nu-2\nu\chi^{2}+3\chi^{2}\right)
\left(b-c\right)}{\left(1-\nu\right)\left(\chi^{2}-\nu\right)}\right],\quad\quad\quad\quad\quad\quad\quad\quad\quad\quad\quad\quad\quad\quad\nonumber
\end{eqnarray}
\begin{eqnarray}
N_{3}=\frac{3}{8}\frac{\mu\left(b-c\right)\left(\nu^{2}-2\nu-2\nu\chi^{2}+3\chi^{2}\right)^{2}
}{\left(1-\nu\right)\left(\chi^{2}-\nu\right)}+\frac{\left(1-\nu\right)
\left(\chi^{2}-\nu\right)}{\left(b-c\right)} \left(4cm
+3c^{2}\mu-\frac{E^{2}-m^{2}}{\mu}\right)+\nonumber\\
+\left(m+\frac{3}{2}c\mu\right)\left(\nu^{2}-2\nu-2\nu\chi^{2}+3\chi^{2}\right)
-\frac{\mu}{2}\left(b-c\right)\left(3\chi^{2}-\left(\chi^{2}+1\right)\nu\right),\quad\quad\quad\quad\quad\quad\nonumber
\end{eqnarray}
\begin{eqnarray}
N_{4}=-\frac{\left(1-\nu\right)
\left(\chi^{2}-\nu\right)}{\left(b-c\right)}\frac{\left(k^{2}-\beta^{2}\right)}{bc\mu},
\quad\quad\quad\quad\quad\quad\quad\quad\quad\quad\quad\quad\quad\quad\quad\quad\quad\quad
\quad\quad\quad\quad\quad\quad\nonumber
\end{eqnarray}

\begin{eqnarray}
N_{5}=\frac{A+\frac{E+m}{2\mu}}{\left(b-A\right)\left(A-c\right)},\quad\quad
N_{6}=\frac{B+\frac{E+m}{2\mu}}{\left(b-B\right)\left(B-c\right)},\quad\quad
N_{7}=\frac{c+\frac{E+m}{2\mu}}{\left(A-c\right)\left(B-c\right)},\quad\quad\nonumber\\
\chi=\sqrt{\frac{\left(a-b\right)\left(c-d\right)}{\left(a-c\right)\left(b-d\right)}},\quad\quad
\nu=\frac{a-b}{a-c},\quad\quad
\nu_{1}=\frac{A-c}{A-b}\nu,\quad\quad
\nu_{2}=\frac{B-c}{B-b}\nu,\quad\quad\quad\quad\quad\quad\nonumber
\end{eqnarray}
Equation (\ref{eq44}) is an implicit relation for the energy
$E_{n k}$.

Accordingly wave functions of light quarks in classically allowed
region are defined by formulae (\ref{eq20}), where normalization
constant is of the form
\begin{equation}\label{eq45}
\left|C_{1}^{\pm}\right|=\left[\frac{\mu\sqrt{\left(a-c\right)\left(b-d\right)}}{2\left[
\left(Ec+\beta\right)\cdot
F\left(\chi\right)+E\left(b-c\right)\cdot\Pi\left(\nu,\chi\right)
\right]}\right]^{1/2}
\end{equation}
and phases $\Theta_{1}$ and $\Theta_{2}$ are
\begin{eqnarray}
\Theta_{1}=-\frac{2}{\sqrt{\left(a-c\right)\left(b-d\right)}}\left[
\frac{\left(b-c\right)^{2}}{\left(1-\nu\right)\left(\chi^{2}-\nu\right)}\times\quad\quad\quad\quad\right.\nonumber\\
\times\left[N_{1}\cdot F\left(\phi,\chi\right)+N_{2}\cdot
E\left(\phi,\chi\right)+ N_{3}\cdot
\Pi\left(\phi,\nu,\chi\right)+N_{4}\cdot
\Pi\left(\phi,\frac{c}{b}\nu,\chi\right)+L\right]
+\label{eq46}\\
\left.+ \frac{k}{\mu}\left[\frac{b-c}{A-B}\left(N_{5}\cdot
\Pi\left(\phi,\nu_{1},\chi\right)-N_{6}\cdot
\Pi\left(\phi,\nu_{2},\chi\right)\right)+N_{7}\cdot
F\left(\phi,\chi\right)\right]\right]+\frac{\pi}{4},\nonumber
\end{eqnarray}
\begin{eqnarray}
\Theta_{2}=-\frac{2}{\sqrt{\left(a-c\right)\left(b-d\right)}}\left[
\frac{\left(b-c\right)^{2}}{\left(1-\nu\right)\left(\chi^{2}-\nu\right)}\times\quad\quad\quad\quad\right.\nonumber\\
\times\left[N_{1}\cdot F\left(\phi,\chi\right)+N_{2}\cdot
E\left(\phi,\chi\right)+ N_{3}\cdot
\Pi\left(\phi,\nu,\chi\right)+N_{4}\cdot
\Pi\left(\phi,\frac{c}{b}\nu,\chi\right)+L\right]
-\label{eq47}\\
\left.- \frac{k}{\mu}\left[\frac{b-c}{A_{1}-B_{1}}\left(N_{8}\cdot
\Pi\left(\phi,\nu_{1}^{\ast},\chi\right)-N_{9}\cdot
\Pi\left(\phi,\nu_{2}^{\ast},\chi\right)\right)+N_{10}\cdot
F\left(\phi,\chi\right)\right]\right]+\frac{\pi}{4}.\nonumber
\end{eqnarray}
Here \[\phi=\arcsin
\sqrt{\frac{\left(a-c\right)\left(r-b\right)}{\left(a-b\right)\left(r-c\right)}},\]
$A_{1}$, $B_{1}$ are roots of the quadratic equation $\mu
r^{2}-\left(E-m\right)r-\beta=0$,
\begin{eqnarray}
L=\frac{\sin\phi\cdot\cos\phi\cdot\sqrt{1-\chi^{2}\sin^{2}\phi}}{1-\nu\sin^{2}\phi}\times
\quad\quad \quad\quad\quad\quad \quad\quad\nonumber\\
\times\left[\frac{3\left(b-c\right)\left(\nu^{2}-2\nu-2\nu\chi^{2}+3\chi^{2}\right)^{2}}
{8\left(1-\nu\right)\left(\chi^{2}-\nu\right)}+\frac{1}{4\left(1-\nu\sin^{2}\phi\right)}+
\frac{m}{\mu\left(b-c\right)}\right],\nonumber
\end{eqnarray}
\begin{eqnarray}
N_{8}=\frac{A_{1}-\frac{E-m}{2\mu}}{\left(b-A_{1}\right)\left(A_{1}-c\right)},\quad\quad
N_{9}=\frac{B_{1}-\frac{E-m}{2\mu}}{\left(b-B_{1}\right)\left(B_{1}-c\right)},\quad\quad
N_{10}=\frac{c-\frac{E-m}{2\mu}}{\left(A_{1}-c\right)\left(B_{1}-c\right)},\quad\quad\nonumber\\
\nu_{1}^{\ast}=\frac{A_{1}-c}{A_{1}-b}\nu,\quad\quad \quad\quad
\nu_{2}^{\ast}=\frac{B_{1}-c}{B_{1}-b}\nu.\quad\quad\quad\quad
\quad\quad\quad\quad\quad\quad \quad\quad\nonumber
\end{eqnarray}

\vspace{0.5cm}\noindent{\Large\bf 6. Quasi-independent quark model
of hadrons}\vspace{0.5cm}

\noindent One of simple of meson quarks models, which, despite of
singleness, gives the good agreement with experimental data, is
the quasi-independent quark model \cite{Bog, Khr}. Instead of to
consider a problem on the bound states of system of two quarks,
intensive interacting one another, we shall consider quarks in a
meson as independent particles, which move in mean or
self-consistent field.

In such scheme not only wavefunction of all system of two quarks
has a sense, but also individual one-particle wavefunctions of
each quark. These one-particle wavefunctions are described by the
usual Dirac equation for particle in some field. Thus, quarks,
which are included in a structure of a meson, occupy one-particle
energy levels. To simplify calculations it is assumed that this
mean field is spherically symmetric and quark motion in space is
determined by motion of its center. Each of constituents
interacting with the spherical mean field gets the state with a
definite value of its energy. On the phenomenological ground the
mass formula for $J^{PC}$ meson, which is $n^{2S+1}L_{J}$ state of
$qq$-system, can be represented in the following form:
\begin{equation}\label{eq48}
M=\sqrt{p^{2}_{1}+m^{2}_{1}}+\sqrt{p^{2}_{2}+m^{2}_{2}}.
\end{equation}
In a system of the centre of mass
($\left|p_{1}\right|=\left|p_{2}\right|=p$) after simple
mathematical transformation we have obtained the relation
\begin{equation}\label{eq49}
\frac{M^{2}+m^{2}_{1}-m^{2}_{2}}{2M}=\widehat{\alpha}p+\beta
m_{1}=E,
\end{equation}
from which
\begin{equation}\label{eq50}
M=E+\sqrt{E^{2}-m^{2}_{1}+m^{2}_{2}}.
\end{equation}
Here $E$ is the eigenvalue of the one-particle Dirac equation
(\ref{eq2}).

\vspace{0.5cm}\noindent{\Large\bf 7. Results and
conclusions}\vspace{0.5cm}

\noindent The Dirac equation gives the spin-orbit splitting only
into two levels while the experiment gives three levels. It is
due to the fact that the individual quark spins can be composed
in 0 or 1 and the total angular moment can be 0, 1 or 2 in the
case $l=1$. But in the limit of an infinitely heavy quark mass
$m_{2}\rightarrow \infty$ degrees of freedom of quarks system are
determined by the quantum states of the light quark with the
total angular moment $j=L+s$. This gives two sets of levels
$j=1/2$, $j=3/2$. For example, in the hydrogen atom, we don't
take into account the nuclear spin: it is entered only as a
hyperfine effect. For comparing our data with other data of
therms with total moment we hold the following rule: $P_{1/2}$
was considered as being an averaged mixture of $^{1}P_{1}$ and
$^{3}P_{0}$, while $P_{3/2}$ was considered as averaged mixture
of $^{3}P_{1}$ and $^{3}P_{2}$ masses. This problem can have
different solutions and it will be discussed in our next papers.

Everyone dealing with the Dirac equation encounters the problem
of Lorenz structure of the potential. Our calculations
demonstrate that considered one-gluon potential being totally
vector and the confinement potential being totally scalar lead to
incorrect spin-orbit splitting of $P$-levels:
$M(P_{3/2})<M(P_{1/2})$. For correct description of sign of
spin-orbit splitting in the Dirac equation it is necessary to
take a part of confinement potential as vector that corresponds
to (\ref{eq36}). Hereinafter we will consider the case
$\varepsilon \neq 0$. This conclusion concerning the Lorenz
character of the confinement agrees with other authors \cite{Cra,
Hay}.

\end{document}